\documentclass[11pt]{article}
\usepackage{graphics,epsfig}
\usepackage{amssymb}
\begin{document}
\setlength{\parindent}{0pt}
\linespread{1.3}
\hyphenation{co-e-ffi-cients Se-ve-ral processing}
\title{{ \bf CONSTRUCTION OF AN ALGORITHM IN PARALLEL FOR THE
FAST FOURIER TRANSFORM}}
\author{Higuera G. Mario A.,
 Sarria Humberto,  
 Fonseca Diana,  
 Idarraga John $^*$} 
\maketitle

\begin{abstract}
It has been designed, built and executed a code for the Fast
Fourier Transform (FFT), compiled and executed in a cluster of $ 2^ n$ computers under the operating system
MacOS and using the routines MacMPI. As practical application, the code has
been used to obtain the transformed from an astronomic imagen, to execute a filter on its and
with a transformed inverse to recover the image with the variates given by the filter.  The computers arrangement 
are installed in the Observatorio Astron\'omico National in Colombia under the name{\it OAN
Cluster}, and in this has been executed several applications.


\end{abstract}

{\tiny { \ }}\\
\line(1,0){100}\\
Observatorio Astron\'omico Nacional\\
Facultad de Ciencias. Universidad Nacional de Colombia.\\ 
e-mail: $^*$ ahiguera@ciencias.ciencias.unal.edu.co, 
hsarria@ciencias.ciencias.unal.edu.co,  dfonseca@ciencias.ciencias.unal.edu.co,
 idarraga@ciencias.ciencias.unal.edu.co  \\ Apartado A\'ereo 2584.
Bogot\'a, Colombia.\\

\section{Introduction}

In the last few years, the implementation of low-cost computer nets for processing in
parallel has been growing. Most of the designated implementations denominate ``Clusters" have been
developed in work stations under environments of Unix; however today exits Clusters that include the
processing G3-G4 of Motorola-IBM-Apple, in computers Apple Macintosh running under MacOS,\cite{appleseed}
and also with Pentium III under Linux or Windows.\cite{pccluster}\\

The{\it OAN Cluster} is a project that born in December of 1998 motivated by the reached developments
for Decyk and his team \cite{decyk}, by the hardware (Macs/G3) recently incorporate to the Observatorio
Astron\'omico Nacional (OAN) and by the developments in software carried out within the investigation lines: 
Theoretical Astrophysics, Galactic Astronomy and Numerical Methods of the OAN. With the support of Absoft
Corporation \cite{absort} was incorporated an excellent software compiler (Fortran 77/90 and C, C ++); that
together with the routines MacMPI developed by Decyk in \cite{internet} completed the development structure of
the OAN Cluster.\\

The first codes were centered in the knowledge of the own commands of the MPI, to evolve
in the data distribution and operations with matrix. Several applications have been
developed, one of them the denominated ``prime numbers under the rule of Stanislav Ulam" the one which
is found available in the Web address of the OAN \cite{oan} and an algorithm in parallel for
the Fast Fourier Transform.\\

\section{Fast Fourier Transform}
Let $f :\Bbb R \to \Bbb C$ a continuously function, where $R$ represents the set of real numbers
and $C$ the complex numbers. The {\it Fourier Transform} of
$f$ is given by
\begin{equation} \label{E1}
H(\nu) = \int_{-\infty}^{\infty}f(t)e^{- 2\pi i\nu t} dt, 
\end{equation}
where $i$ is the imaginary unity and $\nu$ the frecuency \cite{numerical}.\\

In most of practical situations, the function $f$ is given in discrete form as a
finite values collection $f(x_0), f(x_1), \dots ,f(x_{N-1})$ with $N \in \Bbb
N$, where  $\Bbb N$ is the set of natural numbers and 
\{$x_0, x_1, \dots, x_{N-1}$\} is one partition on an real interval $[a, b]$ and $x_k = a+\frac{b-a}{N}k$, for
$0 \leq k \leq N-1$. In problems that imply numerical calculation, instead the equation (\ref{E1}) we use
the sum partial

\begin{equation} \label{E2}
H(k)={1\over N} \sum_{j=0}^{N-1} f(x_j)e^{- 2\pi ikj/N},
\end{equation}
designated  {\it Discrete Fourier Transform} (DFT) of $f$ over the interval $[a, b]$. If $f$ is a function
defined on the interval $[0,2\pi]$ of real value with period
$2\pi$,  the values $H(k)$, by 
$k=0,\dots , N-1,$ can be interpreted as the coefficients $c_k$ 
\begin{equation} \label{coefFourier}
c_k=\frac{1}{N}\sum_{j=0}^{N-1}f(x_j)e^{-i\frac{2\pi k}{N}j}, \qquad \hbox{$0 \leq k \leq N-1$
} 
\end{equation}
of a  {\it exponential polynomial} 
\begin{equation} \label{E3}
p(x) = \sum_{k=0}^{N-1} c_k e^{ikx}, 
\end{equation}
where $x\in \Bbb R$,  $c_k \in \Bbb C,$ y $k=0,1,\dots, N-1,$  which interpole
$f$ in the values 
$f(x_0),f(x_1),\ldots,f(x_{N-1}).$\\ 

The Discrete Fourier Transform of $f$ over the  partition $\{x_0, x_1, \dots, x_{N-1}\},$
is defined as the operator $$DFT: \Bbb C^N\longrightarrow \Bbb C^N$$ such that
\begin{center}
$[c_0, c_1,\ldots, c_{N-1}]^T=DFT\Big([f(x_0), f(x_1),\ldots,f(x_{N-1})]^T\Big)$
\end{center}

The algorithm that evaluates  DFT is denominated Fast Fourier
Transform, and reduces the calculation significantly. For directly
calculations is necesary to use $N^2$
multiplications, while FFT only needes
$NLog_2N$ products. A comparison for greater values of $N$ is shown in
Table \ref{tabla1}.\\

The next tree show how is running the interpolation. In the first level
(indicated by the superscript of the polynomials) are related two
data in each branch that comes of the zero level and produce a
polynomial with two coefficients
$P_k^{(1)},$ for $0\le k\le 3$. In the second level, again the
points gene-\ rated are interpolated and produces new factors and
polynomials
$P_k^{(2)},$ with
$0\le k\le 1$; Finally we obtain the output vector, where
$P_0^{(3)}$ is the polynomial that we requires .\\

\label{arbol}
\hspace{5.8cm}$P_0^{(3)}$\\
\hspace*{5.0cm}$\nearrow$ \hspace{1.5cm}$\nwarrow$\\
\hspace*{3.0cm}$P_0^{(2)}$  \hspace{5.0cm}$P_1^{(2)}$\\
\hspace*{3.0cm}$\nearrow\nwarrow $  \hspace{5.0cm}$ \nearrow\nwarrow$\\
\hspace*{1.5cm}$P_0^{(1)}$  \hspace{2.0cm}$P_2^{(1)}$ 
\hspace{2.5cm}$P_1^{(1)}$ \hspace{2.0cm}$P_3^{(1)}$\\
\hspace*{1.5cm}$ \nearrow\nwarrow$  \hspace{2.0cm}$\nearrow\nwarrow $ 
\hspace{2.5cm}$\nearrow\nwarrow $ \hspace{2.0cm} $\nearrow\nwarrow $\\
\hspace*{0.5cm}$P_0^{(0)}$ \hspace{1.0cm}$P_4^{(0)}$ 
\hspace{0.4cm} $P_2^{(0)}$ \hspace{0.7cm}$P_6^{(0)}$ 
\hspace{1.0cm}$P_1^{(0)}$  \hspace{1.1cm}$P_5^{(0)}$ 
\hspace{0.4cm}$P_3^{(0)}$ \hspace{1.0cm}$P_7^{(0)}$\\

The parallel code works $2^l$
processors,
$l\ge 0$. For a better comprehension let us consider an example: let
$C=[C(0),\ldots,C(7)]$ eight data and four processors.\\

The basic variables that are used, are presented in Table
\ref{tabla2}. Then, for this example we have: $nproc=4$,
$N=8$,
$m=3$,
$m1=1$.
\\  

\begin{table}[h]
\caption{Basic variables}\label{tabla2} 
\begin{center} 
\begin{tabular}{cl}
\hline 
{ \it Variable} & { \it Description} \\
\hline
$nproc$ & Number of processors.\\
$idproc$ &Label correspondent to each processor.\\
$ $ &  We have $0\le idproc< nproc$, with the processor $0$ equal to MASTER.\\
$N$ & number of data.\\
$C$ & Vector with the data. \\
$Z$ & Vector with factors $e^{-\frac{2\pi i}{N}j}$,  $0\le
j\le N-1$.\\
$m$ & Number of tree's levels calculated for
$\Big[\frac{Log(N)}{Log(2)}\Big]$.\\
$n$ & Number of level in the process, $0\le n < m$.\\ 
$m1$ & Number of level until work all processors, $m-\Big[\frac{Log(nproc)}{Log(2)}\Big]$. \\
$nproc\_sup$ & Number of processors in some levels.\\
\hline 
\end{tabular} 
\end{center} 
\end{table}

Each processor works with its respective  branch:\\

\hspace{1.1cm}$idproc=0$\hspace{1.2cm}$idproc=2$ 
\hspace{1.5cm}$idproc=1$\hspace{1.1cm}$idproc=3$\\
\hspace*{1.8cm}$P_0^{(1)}$  \hspace{2.0cm}$P_2^{(1)}$ 
\hspace{2.6cm}$P_1^{(1)}$ \hspace{2.2cm}$P_3^{(1)}$\\
\hspace*{1.7cm}$\nearrow\nwarrow $  \hspace{2.0cm}$ \nearrow\nwarrow$ 
\hspace{2.5cm}$ \nearrow\nwarrow$ \hspace{2.0cm} $\nearrow\nwarrow $\\
\hspace*{0.7cm}$P_0^{(0)}$ \hspace{1.0cm}$P_4^{(0)}$ 
\hspace{0.4cm} $P_2^{(0)}$ \hspace{0.7cm}$P_6^{(0)}$ 
\hspace{1.0cm}$P_1^{(0)}$  \hspace{1.1cm}$P_5^{(0)}$ 
\hspace{0.4cm}$P_3^{(0)}$ \hspace{1.0cm}$P_7^{(0)}$\\
\hspace*{0.6cm}$C(0)$\hspace{1.0cm}$C(4)$ 
\hspace{0.3cm} $C(2)$\hspace{0.7cm}$C(6)$ 
\hspace{0.8cm}$C(1)$\hspace{0.9cm}$C(5)$ 
\hspace{0.4cm}$C(3)$\hspace{0.9cm}$C(7)$\\

In this step, the processor identified with the label $idproc$ generates the
coefficients' vector:
$P^{(n)}_{idproc}$ for
$0\leq idproc < nproc=4;$ ($n=1$). The processing of the information continues until 
$n=m1=1$. In this point $idproc=2$ transmits its vector to $idproc=0,$ and $idproc=3$
transfers its vector to $idproc=1$. Before of the transference, each processor
stores the information in a matrix with two columns
where the real part is stored in the first column and the imaginary part in the
second column. This is necessary because the commando to send don't recognize
complex numbers in language C.\\ 

In this moment $idproc=2$ and $idproc=3$ don't work. Moreover, 
$nproc\_sup=\frac{nproc}{2}=2,$ so, in the level $n=2$ the number of  processors is
reduced to the half. From now on, where we avanced of level, the middle of processors out of the game and the
variable
$nproc\_sup$ descend to the middle.\\ 

The processors that receive the information unpack in $C$.\\

\hspace{2.5cm}$idproc=0$\hspace{4.0cm}$idproc=1$\\
\hspace*{3.2cm}$P_0^{(2)}$  \hspace{5.0cm}$P_1^{(2)}$\\
\hspace*{2.3cm}$C(0)$\hspace{0.5cm}$C(2)$\hspace{3.8cm}$C(4)$\hspace{0.5cm}$C(6)$\\
\hspace*{2.3cm}$C(1)$\hspace{0.5cm}$C(3)$\hspace{3.8cm}$C(5)$\hspace{0.5cm}$C(7)$\\
\hspace*{3.0cm}$\nearrow\nwarrow $  \hspace{5.0cm}$\nearrow\nwarrow $\\
\hspace*{2.2cm}$P_0^{(1)}$  \hspace{0.7cm}$P_2^{(1)}$ 
\hspace{3.5cm}$P_1^{(1)}$ \hspace{0.9cm}$P_3^{(1)}$\\

In the level $n=2$, $idproc$ bring about the vector of coefficients 
$P^{(n)}_{idproc}$ for $0\leq idproc < nproc\_sup=2$. That which is stated previously, is repeated while
$nproc\_sup\ge 1$, that is to say, while exist one processor. The MASTER receive
the communication of $idproc=1$ and effects the last calculation to bring out
$P^{(n)}_{idproc}$ ($idproc=0$ and $n=m=3$), in this instant
$nproc\_sup=\frac{1}{2}$ and the process finish.

\section{THE INVERSE TRANSFORM}

We affirm that the exponential polinomyal (\ref{E3})
estimates the Inverse Discrete Fourier Transform (IDFT), definied as the operator
$$IDFT: \Bbb C^N\longrightarrow \Bbb C^N$$ such that

\begin{center}

$[f(x_0), f(x_1),\ldots,f(x_{N-1})]^T=IDFT\Big([c_0, c_1,\ldots,c_{N-1}]^T\Big)$

\end{center}

We use the FFT to calculate the IFFT\footnote{Inverse Fast Fourier Transform}. We define $x_k=\frac{2\pi
k}{N}$ for \\$0\leq k\leq N-1$, and evaluate $p$ in $2\pi -x_k=2\pi
-\frac{2\pi k}{N}$

\begin{equation}
p(2\pi-x_k)=\sum_{j=0}^{N-1}c_je^{ij(2\pi-x_k)}=
\sum_{j=0}^{N-1}c_je^{ij2\pi}c_je^{-ijx_k}=
\sum_{j=0}^{N-1}c_je^{-ijx_k}= N\Big(\frac{1}{N}\sum_{j=0}^{N-1}c_je^{-ijx_k}\Big)
\end{equation}

In this way obtain the IFFT estimating the FFT multiplied for $N$ and 
$$p(2\pi-x_k)=p(2\pi-\frac{2\pi k}{N})=p\Big(2\pi \frac{N-k}{N}\Big),$$ so the inverse transform
is in contrary order.\\

\section{COMPLEXITY COMPUTATIONAL OF FFT IN PARALLEL}

If the number of data is $N=2^m$ ($m\geq 1$) and the number of processors is $2^l$ ($l\geq 0$),
the order of complex multiplications effected until the level $m1$, correspond to the  multiplications made for
each processor with its respective branch. The number of  multiplications carried out until the level 
$n$ is $n2^m$ for
$0\leq n\leq m.$ Particulary, if $n=m1$ we have $O(m1(2^m))$. Then dividing for the number of processors, the
complexity descend to
$$\frac{m1(2^m)}{2^l}=(m1)2^{m-l}=m1(2^{m1})$$ for
$n=m1$.\\

Finally we adds the number of multiplications effected in the levels
$n$, for $m1 < n\leq m.$ In each level we have $2^m$ products. Considering that for each level $n$, $m1 < n\leq
m,$ the number of processors is reduced to the middle, obtain $$\sum_{i=1}^l2^{m1+i}$$
from $m1+1$ until $n=m$.\\ 
Summarizing,  
\begin{equation}\label{multi}
O\Big((m-l)2^{m-l}+\sum_{i=1}^l2^{m-l+i}\Big).
\end{equation}
Calculating the ratio between sequential FFT and parallel FFT, we have
\begin{eqnarray}
\frac{NLog_2N}{(m-l)2^{m-l}+\sum_{i=1}^{l}2^{m-l+i}}&=&\frac{m2^m}{(m-l)2^{m-l}+\sum_{i=1}^{l}2^{m-l+i}}\nonumber\\
&=&\frac{m}{(m-l)2^{-l}+\sum_{i=1}^{l}2^{-l+i}}\nonumber
\end{eqnarray}
This indicate that the algorithm in parallel is $\frac{m}{(m-l)2^{-l}+\sum_{i=1}^{l}2^{-l+i}}$ faster than the
sequential algorithm. 

\begin{table}[h]
\caption{Comparative arithmetic products} \label{tabla1}
\begin{center}
\vspace{1mm}   
\begin{tabular}{ccccc}
\hline
$N$ & $N^2$ & $NLog_2N$ & \emph{Pfp}
& \emph{RSP FFT}\\
\hline
512 & 262144 & 4608 & 1664 & 2.77\\
2048 & 4194304 & 22528 & 7680 & 2.93\\
8192 & 67108864 & 106496 & 34816 & 3.06\\
32768 & 1073741824 & 491520 & 155648 & 3.16\\
\hline 
\end{tabular} 
\end{center} 
{\it Pfp}: Paralell with four processors.\\
{\it RSP FFT}: Ratio between sequential and paralell FFT.
\end{table}

\section{THE TWO-DIMENSIONAL DISCRETE TRANSFORM 2D-DFT}

For two variables the sample is in $xy$-plane where the sample is uniformly distribuited in the
parallel straight-lines
to the $x$-axis (rows) and the parallel straight-lines
$y$-axis (columns). We define $\mathcal{M}_{N\times M}(\Bbb C)$ the matrix of size
$N\times M$ which contain complex numbers. Let $f(x,y): \Bbb R^2\longrightarrow \Bbb C$ be a
two-dimensional function, its 2D-DFT is definied as the operator
$$2D-DFT: \mathcal{M}_{N\times M}(\Bbb C)\longrightarrow \mathcal{M}_{N\times M}(\Bbb C)$$
\hspace{5.0cm}$\Big(f(x,y)\Big)_{N\times M}\mapsto \Big(F(u,v)\Big)_{N\times M}$\\
where
\begin{equation}
\label{eq5}
F(u,v)=\frac{1}{N}\sum_{x=0}^{N-1}\sum_{y=0}^{M-1}f(x,y)e^{-i2\pi(\frac{ux}{N}+\frac{vy}{M})}
\end{equation}

for $u=0, 1,\ldots,N-1, v=0, 1,\ldots,M-1.$\\
And we establish the 2D-IDFT  (Inverse Discrete Fourier Transform
two-dimensional) as $$2D-IDFT:
\mathcal{M}_{N\times M}(\Bbb C)\longrightarrow \mathcal{M}_{N\times M}(\Bbb C)$$
\hspace{5.0cm}$\Big(F(u,v)\Big)_{N\times M}\mapsto \Big(f(x,y)\Big)_{N\times M}$\\
where
\begin{equation}
f(x,y)=\frac{1}{M}\sum_{u=0}^{N-1}\sum_{v=0}^{M-1}F(u,v)e^{i2\pi(\frac{ux}{N}+\frac{vy}{M})}
\end{equation}

for $x=0, 1,\ldots,N-1, y=0, 1,\ldots,M-1.$\\

The equation (\ref{eq5}) we express as
\begin{equation}
F(u,v)=\frac{1}{N}\sum_{x=0}^{N-1}\sum_{y=0}^{M-1}f(x,y)e^{-i2\pi\frac{ux}{N}}e^{-i2\pi\frac{vy}{M}}
=\frac{1}{N}\sum_{x=0}^{N-1}F(x,v)e^{-i2\pi\frac{ux}{N}}
\end{equation}

where $F(x,v)=M\Big(\frac{1}{M}\sum_{y=0}^{M-1}f(x,y)e^{-i2\pi\frac{vy}{M}}\Big)$. In this way, we may
frequent use of the DFT on rows and on columns.\\ 

The implementation of the algorithm for the 2D-FFT (Fast Fourier Transform two-dimensional)
is, 
\\

FFT on rows $\longrightarrow$ Multiplication for $M$ $\longrightarrow$ FFT on
columns. \cite{imagen}\\

\section{APPLICATION}

A application has been diriged to the image processing. We can represent a image
as a function
\begin{center}
$f: \Bbb R^2\longrightarrow \Bbb Z$\\
\begin{tabular}{c c c c c c}
& &  & & & $(x,y)\mapsto z=f(x,y)$\\
\end{tabular}\\
\end{center}
where $\Bbb Z$ is the set of integer numbers, $(x,y)$ is the coordinate image's
point with brightness
$f(x,y)$. A digital image, it's a image which $x, y,  f(x,y) \in \Bbb Z^+\cup\{0\}$. Thus, a image
is a two-dimensional array $(P_{x,y})_{N\times M}$ of pixels (picture elements).\\

The algorithm pads with zeros on rows and on columns to complete the next power
of two.\\

The Fourier Transform we can represent trought of its spectrum
\begin{equation}
\sqrt{Re[C(x,y)]^2+Im[C(x,y)]^2}
\end{equation}

where $Re[C(x,y)]$ is the real part and $Im[C(x,y)]$ is the imaginary part of the Transform's element
$(x,y)$.\\

A images' filter is a operator $H: \Bbb Z^2\longrightarrow \Bbb R$ which permits
to change the brightness in the digital image. For the Convolution Theorem  we
can use a filter and to multiply it with the real and imaginary part image's
Fourier Transform. A filter can be designed either to eliminate or to create
noise in a image. We use the filter:\\

\begin{equation}
\label{filtro}
F\left( {x,y} \right)=\left\{ {\matrix{{0\ }\cr
{1\ }\cr
}\matrix{{si\ \sqrt {\left( {x-c_1} \right)^2+\left( {y-c_2} \right)^2}\le b}\cr
{en\ caso\ contrario\ \ \ \ \ \ \ \ \ \ \ \ \ \ \ \ }\cr
}} \right.
\end{equation}\\

where $(c_1,c_2)$ is the image's center and $b>0$,  \cite{imagen}.\\

Finally, we calculate the 2D-IFFT to this product and obtain the image
filtered. In the next step, we can see the smoothing effect derived using the filter over the spiral galaxy
NGC5194 (Whirpool Galaxy). This picture has a size of 460$\times$506 pixels, and was taken of
\cite{galaxia}; in our work we use a filter parameter of 
$b=340$.
\\ 

\begin{figure} \label{original} 
\hspace {2.5cm}
\includegraphics[angle=0,width=7cm]{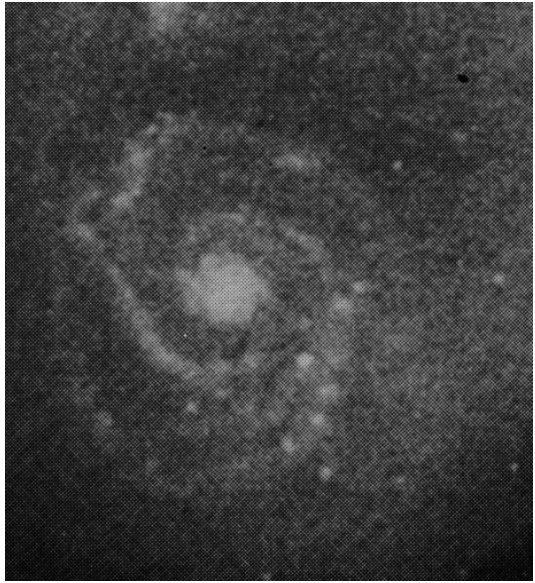}
\caption{Original galactic picture} \label{original}
\end{figure}

\begin{figure} \label{espectro}
\hspace {2.5cm}
\includegraphics[angle=0,width=7cm]{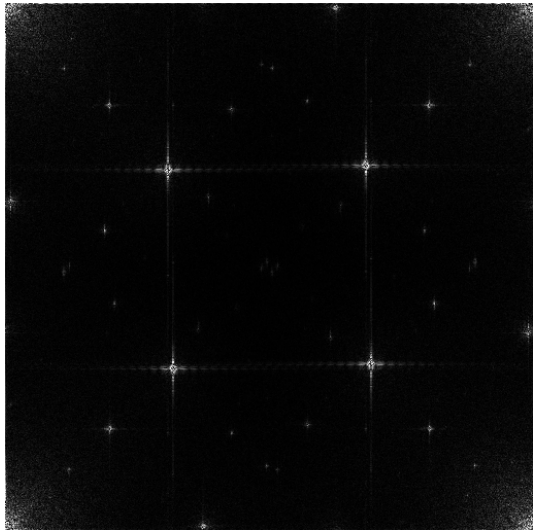}
\caption{Fourier spectrum} \label{espectro}
\end{figure}

\begin{figure}\label{filtra}
\hspace {2.5cm}
\includegraphics[angle=0,width=7cm]{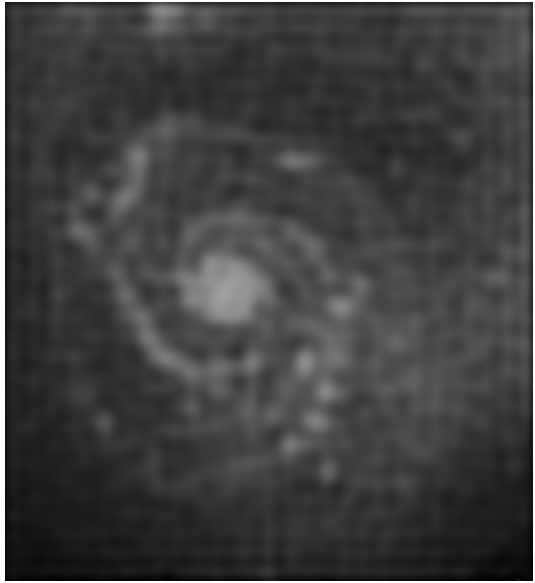}
\caption{Transformed picture } \label{filtra}
\end{figure} 

\section{SUMMARY AND CONCLUSIONS}

\begin{enumerate}
\item Arithmetical expression for number of multiplications in the parallel FFT
was obtained in the equation (\ref{multi}) and calculated the ratio 
between sequential and paralell FFT products (with four processors) for four
values of
$N$, listed in Table
\ref{tabla1}.  

\item We write until the proper functioning the algorithm
2D-FFTp.c. using the MPI's
routines and tools of MacOS system.

\item We find that the efficiency increase when the communications has the
minimum rate of transference executing the code in the cluster. 

\item We build the filter of the equation (\ref{filtro}) and had been applied
to the galaxy NGC5194 obtaining a new image (figure \ref{filtra}) which the high
frequency's components had been eliminated.
\end{enumerate}

We acknowledge an anonymous referee for this helpful coments. We thank Viktor K. Decyk in UCLA for his useful
recommendatios and permanent \  assistance. This work has been supported by DIB of the Universidad Nacional de
Colombia through Proyect DIB-803577.\\

\end{document}